\documentclass[12pt]{article}
\usepackage{osajnl}
\usepackage{overcite}
\usepackage{graphics}
\usepackage{geometry}

\begin{document}

\title{Influence of structure on the optical limiting properties of nanotubes}

\author{N. Izard$^{1,2,*}$, P. Billaud$^{1}$, D. Riehl$^{1}$ and E.
Anglaret$^{2}$}
\affiliation{1: Centre Technique d'Arcueil, D\'el\'egation G\'en\'erale pour l'Armement,
France}
\affiliation{2: Groupe de Dynamique des Phases Condens\'ees, Universit\'e Montpellier
II, France}

\begin{abstract}
We investigate the role of carbon nanotube structure on their optical limiting
properties. Samples of different and well-characterized structural features are
studied by optical limiting and pump-probe experiments.  A clear influence of
the diameter of the nano-object is demonstrated. Indeed, both nucleation and
growth of gas bubbles are expected to be sensitive on diameter.
\end{abstract}


Nowadays, frequency agile pulsed lasers are widespread. Therefore, development
of optical limiting systems, i.e. self-activated protection against potential
laser hazard, is an actual challenge. Several non linear optical materials were
proposed for optical limiting in the past 15 years, such as reverse saturable
absorbers\cite{RSA}, multiphoton absorbers\cite{TPA} and non linear scattering
systems like carbon black suspensions\cite{CBS} and single-wall carbon nanotube
(SWNT) suspensions\cite{SWNT}. Indeed, the association of materials with
complementarity limiting properties, i.e. non linear scattering from SWNT and
multiphoton absorption from organic chromophores, was recently shown to be a
promising approach to achieve performant optical limiting systems featuring
broad temporal and spectral ranges of efficiency\cite{combinaison}. The main
mechanism responsible for the optical limiting properties of carbon nanotubes
suspension is now well known\cite{SWNT}. Heating of the nanotubes leads to
solvent bubbles formation (via heat transfer from nanotubes to solvent) and to
sublimation of carbon nanotubes themselves, inducing efficient non linear
scattering of the incident laser beam. Optimisation of carbon nanotube
suspensions requires a better understanding of relations existing between carbon
nanotube structure and their optical limiting properties.  Indeed, data
currently published are only fragmentary\cite{MWNT_soluble, MWNT_sus, SWNT_B}.
Either materials studied are not well characterized, or studies concern only a
few samples.

In this letter, we report on the optical limiting behavior of model carbon
nanotubes of various structures. Samples were purchased from MER Inc, Nanoledge
and Nanolab Inc. SWNT were produced by the electric arc process while MWNT were
produced by a CVD process. All samples were extensively characterized using
scanning and transmission electron microscopy, X-Ray diffraction, Raman and
optical spectroscopy\cite{procKirchberg}. The mean diameter of SWNT is
1.35~$\pm$~0.15~nm. Nanotubes assemble into crystalline hexagonal bundles, with
a diameter of about 10 to 15~nm. We also worked on MWNT samples of different
length and diameter distributions, as summarized in table~\ref{TailleMWNT}. 

Aqueous suspensions of nanotubes were realised from these raw samples in water,
using 1 wt\% of Sodium Dodecyl Sulfate (SDS), and will be designated thereafter
SWNT-Bundle, MWNT-1\dots~A suspension of individualised SWNT was also prepared
from the suspensions of SWNT bundles, following the procedure described by O.
Connel \textit{et al.} in reference\,\citeonline{tubesIsoles}, and will be
designated SWNT-Individual.  Evidence of SWNT exfoliation was obtained by direct
HRTEM observation, fluorescence and Raman scattering\cite{papierRaman}. At last,
we prepared a suspension of SWNT shortened by an oxidative treatment, designated
SWNT-Short.  We checked by electron microscopy that these tubes were
significantly shorter (below 100 nm) than the unshortened tubes (\textgreater\,1
$\mu$m) and that the diameter of the bundle was not affected (not shown). Linear
optical transmissions of the suspensions were adjusted to 70 \% at 532 nm in 2
mm thick cells (concentration was around 10 mg/l ). Non linear optical
transmittance measurements were performed using a Q-switched, but non injected,
frequency doubled Nd:YAG, with a pulse duration of 15 ns in a F/50 focusing
geometry. Pump-probe experiments were performed using a frequency doubled Nd:YAG
pump laser emitting 4 ns pulses at 532 nm and a 633 nm continuous probe.

\begin{figure}[p]
\centerline{\includegraphics[width=12cm]{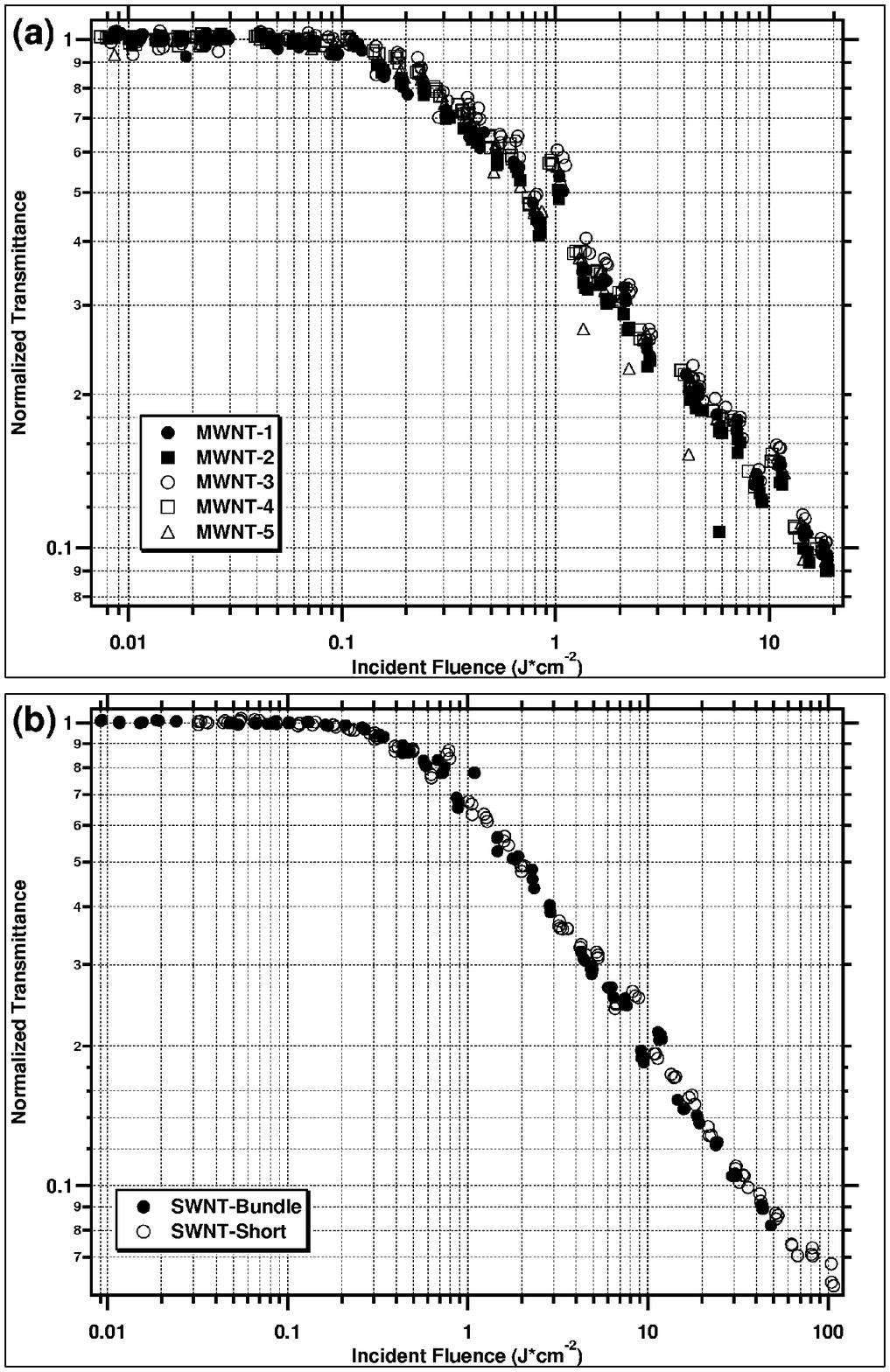}}
\caption{\label{NIfig1}}
\end{figure}

We first investigated the influence of nanotube length. Optical limiting
measurements of the suspensions MWNT-1/5 are reported on figure~\ref{NIfig1}~a.
Optical limiting performances are comparable for all samples, despite the strong
variation in nanotube length (there is a factor 5 between the longest and the
shortest tube). This result clearly states that nanotube length is not a
structural parameter influencing on the optical limiting properties. No diameter
influence was observed neither, but here the diameter variation is small (factor
2 at most). Optical limiting properties of suspensions SWNT-Bundle and
SWNT-Short, which contain nanotubes of same diameter are reported on
figure~\ref{NIfig1}~b. These data confirm the non influence of the length. Note
that Riggs et al.\,\citeonline{riggs} reported slightly weaker optical limiting
performances for shortened nanotubes, but their data cannot assign it to a
length effect : such a behavior could also be due to a slight unbundling of the
nanotubes consecutive to the shortening process, and the lack of pre- and
post-process characterization of nanotube diameter and length do not
allow to choose between the two hypotheses.

\begin{figure}[p]
\centerline{\includegraphics[width=12cm]{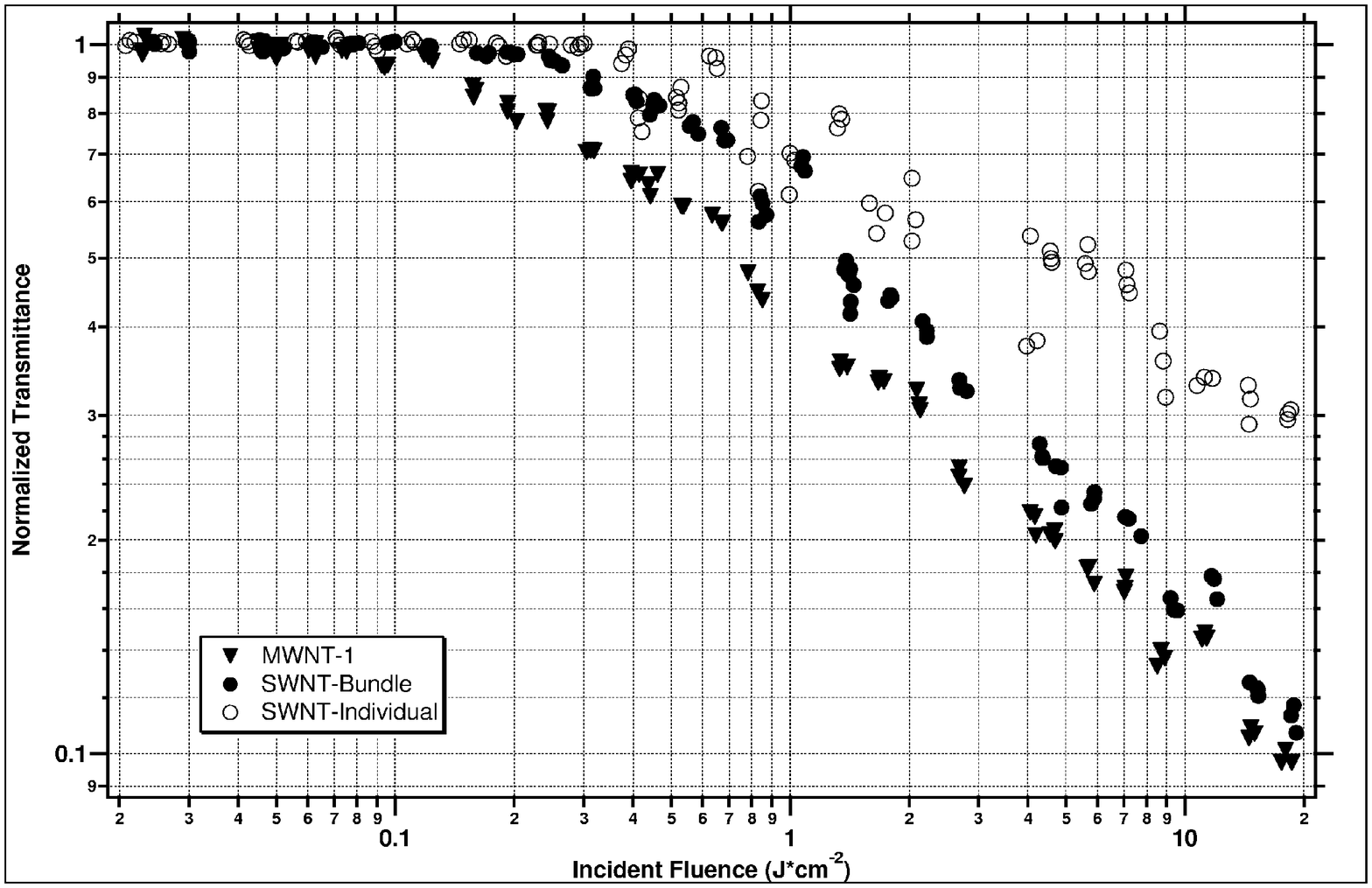}}
\caption{\label{NIfig2}}
\end{figure}

We also studied the influence of nanotube diameter. Figure~\ref{NIfig2}
presents transmittance measurements for suspensions of individual SWNT, bundled
SWNT and MWNT. The diameter of MWNT (20 -- 50 nm) is larger than the diameter of
bundled SWNT (10 -- 15 nm), itself larger than the diameter of individualised
SWNT (1.4 nm). The optical limitation thresholds obtained with MWNT, bundled
SWNT and individual SWNT suspensions are around 100~mJ$\cdot$cm$^{-2}$,
200~mJ$\cdot$cm$^{-2}$ and 400~mJ$\cdot$cm$^{-2}$, respectively. The dependency
of the optical limiting properties on the nano-object diameter is thus clearly
demonstrated. The better efficiency, and the lower optical limiting threshold,
are achieved for the largest diameter of the nano-object.

\begin{figure}[p]
\centerline{\includegraphics[width=12cm]{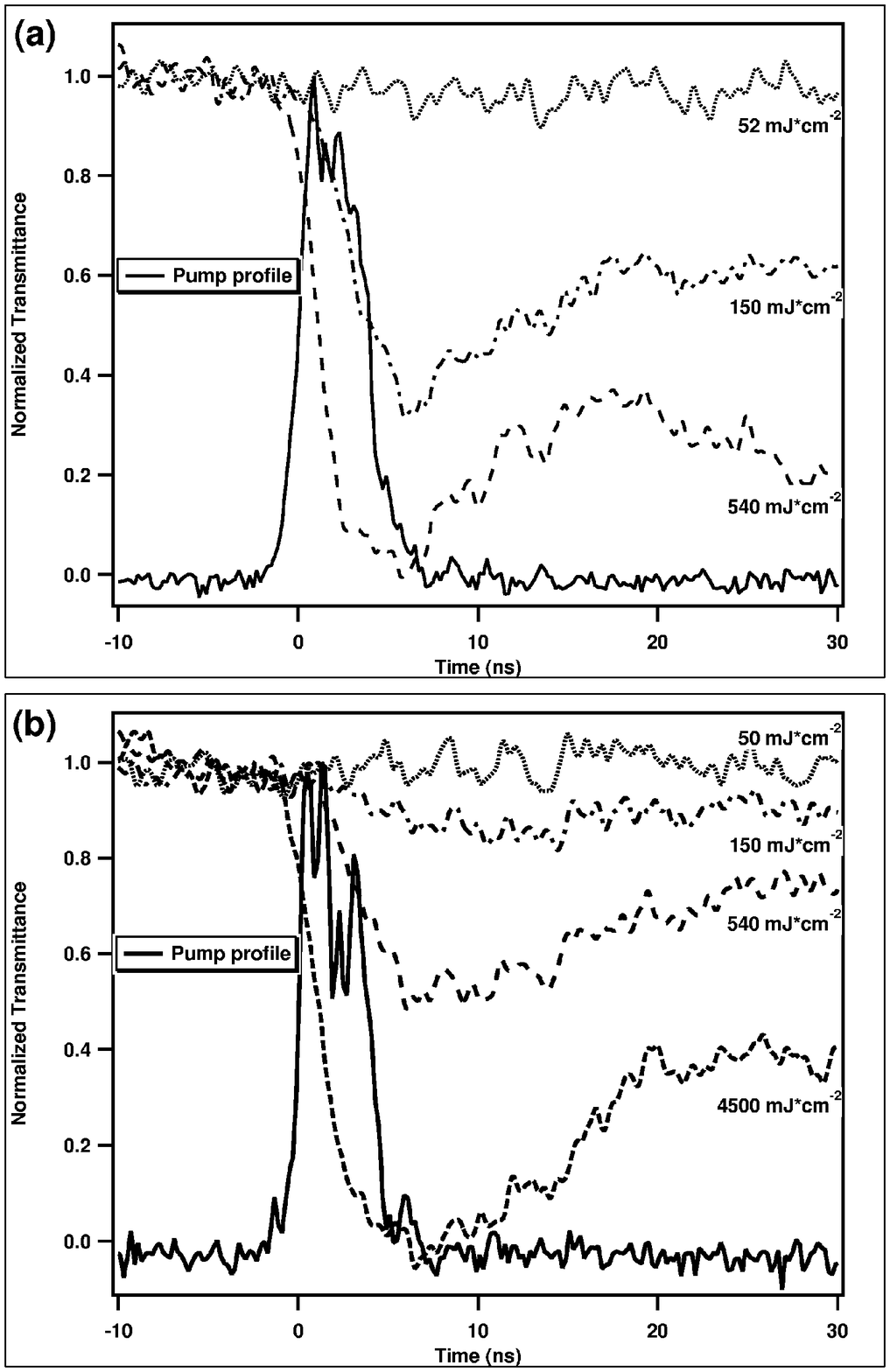}}
\caption{\label{NIfig3}}
\end{figure}

To get a better insight on the effect of the diameter, we carried out
pump-probe experiments on SWNT suspensions, in bundles (Fig.~\ref{NIfig3}~a~)
and individualised (Fig.~\ref{NIfig3}~b~). At low fluence,
52~mJ$\cdot$cm$^{-2}$ for SWNT-Bundle suspension, the probe is not perturbated
by the pump.

The fluence of 150~mJ$\cdot$cm$^{-2}$ corresponds roughly to the limitation
threshold : the probe is perturbed (i.e. bubbles are created) just at the end of
the pump pulse. This threshold value is the limiting threshold, which is
comparable to the value determined from figure~\ref{NIfig2}, although the pulse
is shorter. At last, for higher pump energy, 540~mJ$\cdot$cm$^{-2}$, the
transmission of the probe falls at the beginning of the pump pulse : the
transferred energy is sufficient to sublimate nanotubes and to create bubbles
effective for optical limiting. The behavior is dramatically different for
SWNT-Individual suspensions. Indeed, at 150~mJ$\cdot$cm$^{-2}$, probe
transmission falls slightly more than 10 ns after the passing of the pump. This
means that some bubbles are nucleated inside the suspension by the pump, but
they are too small to be effective for limiting. Fluence must increase up to
540~mJ$\cdot$cm$^{-2}$ to reach the limitation threshold. At
4500~mJ$\cdot$cm$^{-2}$ an efficient optical limiting of the pump beam is
observed. This value is more than 8 times larger for individual tubes than for
bundles.

We attribute this result to the effect of diameter on nucleation and growth of
bubbles. Indeed, it is reasonable to assume than the size of the scattering center
created by the nano-object is a function of its diameter. Consequently,
individualised nanotubes are expected to nucleate smaller bubbles than bundled
nanotubes. This observation leads to two consequencies. From the Laplace law
(here below given for a sphere),

\begin{equation}
P_{ext} = P + \frac{2 \gamma}{R},
\label{laplace}
\end{equation}
where $P$ is the pressure of a gas bubble, $P_{ext}$ the pressure of the
surrounding fluid, $\gamma$ the tensile surface liquid / gas and $R$ the radius
of a bubble, the surpression needed to nucleate a bubble
in a liquid increases when its diameter decreases. So, more energy is involved in the
nucleation of bubbles in individual tube suspensions than for bundled tube
suspensions, i.e.  bundled tube will create bubbles at lower
incident fluence than individuals. Furthermore, once nucleated, the diameter of the bubble is much smaller than
the laser wavelength, so bubbles will scarcely scatter the incident laser beam.
They have to grow until they reach a "critical" (efficient for scattering) size allowing efficient light
scattering. This growth time strongly depends on liquid / gas tensile surface
and other thermodynamical parameters\cite{croissanceBulles}, but also on the
initial bubble size. Thus, a small bubble will take a longer time to reach the
"critical" size than a larger bubble. Therefore,
suspensions of individualised nanotubes will present a poorer optical limiting
efficiency than suspensions of bundled nanotubes.

Finally, it cannot be ruled out that nanotubes in bundles heat faster than
individualised tubes. This will occur if the heat capacity for bundles is
smaller than that for individualised tubes or if absorption cross section per
unit of mass is larger for bundles than for individual. Data on
nanotube thermal properties are scarce and did not make distinction between
tubes in bundles or individualised. Meanwhile, plasmons are
mainly responsible for absorption in carbon nanotubes, and their properties
are expected to be sensitive on environment (including bundling). To go further
in the interpretation, plasmon coupling between carbon nanotubes inside the
bundles may induce a significant absorption enhancement.

In summary, we reported a clear influence of the nanotube diameter on their optical
limiting properties. A two-fold interpretation is proposed. First, a larger
nanotube size involves a larger nucleation center and a faster growth and then
improves optical limiting efficiency. Second, the absorption cross section might
be larger for bundled nanotubes, due to changes in plasmon properties. In conclusion,
increasing nanotube diameter will help improving the optical limiting properties
of suspensions. However, the increase in size must be limited in order to preserve
the stability of suspensions, and to prevent from light scattering at low
fluences. Anyway, this approach is interesting and promising, and is another
step towards the realisation of laser protection for the human eye.

We acknowledge E. Doris and C. M\'enard for fruitful discussions and their
useful advice for the preparation of the samples.

$^{*}$: Email: izard@gdpc.univ-montp2.fr



\newpage

\begin{table}
\begin{center}
\caption{Length and tube diameter distributions for MWNT samples \label{TailleMWNT}}
\begin{tabular}{ccc}
\hline
Sample & Tube diameter (nm) & Length ($\mu$m)\\
\hline
MWNT-1 & 20 -- 50 & 5 -- 20 \\
MWNT-2 & 20 -- 50 & 1 -- 5 \\
MWNT-3 & 10 -- 20 & 5 -- 20 \\
MWNT-4 & 10 -- 20 & 1 -- 5 \\
MWNT-5 & 10 -- 20 & \textless\,1 \\
\hline
\end{tabular}
\end{center}
\end{table}


\newpage

\vskip 1cm

Fig. 1: Normalized transmittance measurements with 15 ns pulses at 532 nm for
(a) MWNT-1 to MWNT-5 suspensions (b) SWNT-Bundle and SWNT-Short suspensions.

\vskip 1cm

Fig. 2: Normalized transmittance measurements with 15 ns pulses at 532 nm for
suspensions SWNT-Individual, SWNT-Bundle and MWNT-1.

\vskip 1cm

Fig. 3: Pump-probe experiments for suspensions of (a) SWNT-Bundle and (b)
SWNT-Individual. In each case, pump profile (532 nm, 4 ns) is represented by a
plain line, while the probe profile (633 nm) is represented by a dashed line for
several pump energies.

\newpage

\clearpage

\clearpage

\clearpage

\end{document}